\documentclass[epj]{webofc}
\usepackage[varg]{txfonts}   
\usepackage{graphicx}
\usepackage{amssymb}
\wocname{EPJ Web of Conferences}
\woctitle{INPC 2013}
\begin{document}
\title{Nuclear Symmetry Energy: constraints from Giant Quadrupole Resonances and Parity Violating Electron Scattering}
%
%

\author{X. Roca-Maza\inst{1,2}\fnsep\thanks{\email{xavier.roca.maza@mi.infn.it}} \and 
B.~K.~Agrawal\inst{3} \and 
P.~F.~Bortignon\inst{1,2} \and
M.~Brenna\inst{1,2} \and
Li-Gang Cao\inst{4} \and
M.~Centelles\inst{5} \and 
G.~Col\`o\inst{1,2} \and 
N.~Paar\inst{6} \and 
X.~Vi\~nas\inst{5} \and  
D.~Vretenar\inst{6} \and   
M.~Warda\inst{7}
}

\institute{
Dipartimento di Fisica, Universit\`a degli Studi di Milano, via Celoria 16, I-20133 Milano, Italy \and
INFN, sezione di Milano, via Celoria 16, I-20133 Milano, Italy \and
Saha Institute of Nuclear Physics, Kolkata 700064, India \and
Institute of Modern Physics, Chinese Academy of Sciences, Lanzhou 730000, China \and
Departament d'Estructura i Constituents de la Mat\`eria and Institut de Ci\`encies del Cosmos, Facultat de F\'{\i}sica, Universitat de Barcelona, Diagonal {\sl 645}, {\sl 08028} Barcelona, Spain \and
Physics Department, Faculty of Science, University of Zagreb, Zagreb, Croatia \and
Katedra Fizyki Teoretycznej, Uniwersytet Marii Curie--Sk\l odowskiej, ul. Radziszewskiego 10, 20-031 Lublin, Poland
}

\abstract{%
Experimental and theoretical efforts are being devoted to the study of observables that can shed light on the properties of the nuclear symmetry energy. We present our new results on the excitation energy \cite{roca-maza13a} and polarizability of the Isovector Giant Quadrupole Resonance (IVGQR), which has been the object of new experimental investigation \cite{henshaw11}. We also present our theoretical analysis on the parity violating asymmetry at the kinematics of the Lead Radius Experiment (PREx \cite{prex}) and highlight its relation with the density dependence of the symmetry energy \cite{roca-maza11}. 
}

\maketitle
\section{Introduction}
\label{intro}
The nuclear symmetry energy is a basic ingredient of the nuclear equation of state: it accounts for the energy cost per nucleon to convert all protons into neutrons in symmetric nuclear matter. Its accurate determination impacts on quantities such as the neutron skin thickness ---defined as the difference between the root mean square radii of neutrons $\langle r_n^2\rangle^{1/2}$ and protons $\langle r_p^2\rangle^{1/2}$ \cite{brown00,tsang12,centelles09}--- and the isovector giant dipole or quadrupole resonances \cite{tsang12,trippa08,roca-maza13a}, amongst others. Isovector collective excitations are characterized by an out of phase oscillation of neutrons against protons, being the restoring force proportional to the nuclear symmetry potential and, therefore, directly related to the symmetry energy. Only recently, the accuracy in the experimental determination of the IVGQR has been substantially improved \cite{henshaw11}.


On the other side, experiments with spin polarized electron scattering by heavy nuclei have been demonstrated to be feasible by the PREx collaboration \cite{prex}. Such experiments can provide a direct and model independent measure of the parity violating asymmetry \cite{donnelly89,horowitz01c}. While the conventional electron scattering experiments determine the electromagnetic distribution in nuclei, the parity violating electron scattering provides unambiguous information on the weak charge distribution in nuclei, basically carried by neutrons \cite{roca-maza11}. 


\section{Giant Quadrupole Resonances}
\label{gqr}

\begin{figure}
\centering
\includegraphics[width=0.45\linewidth,clip=true]{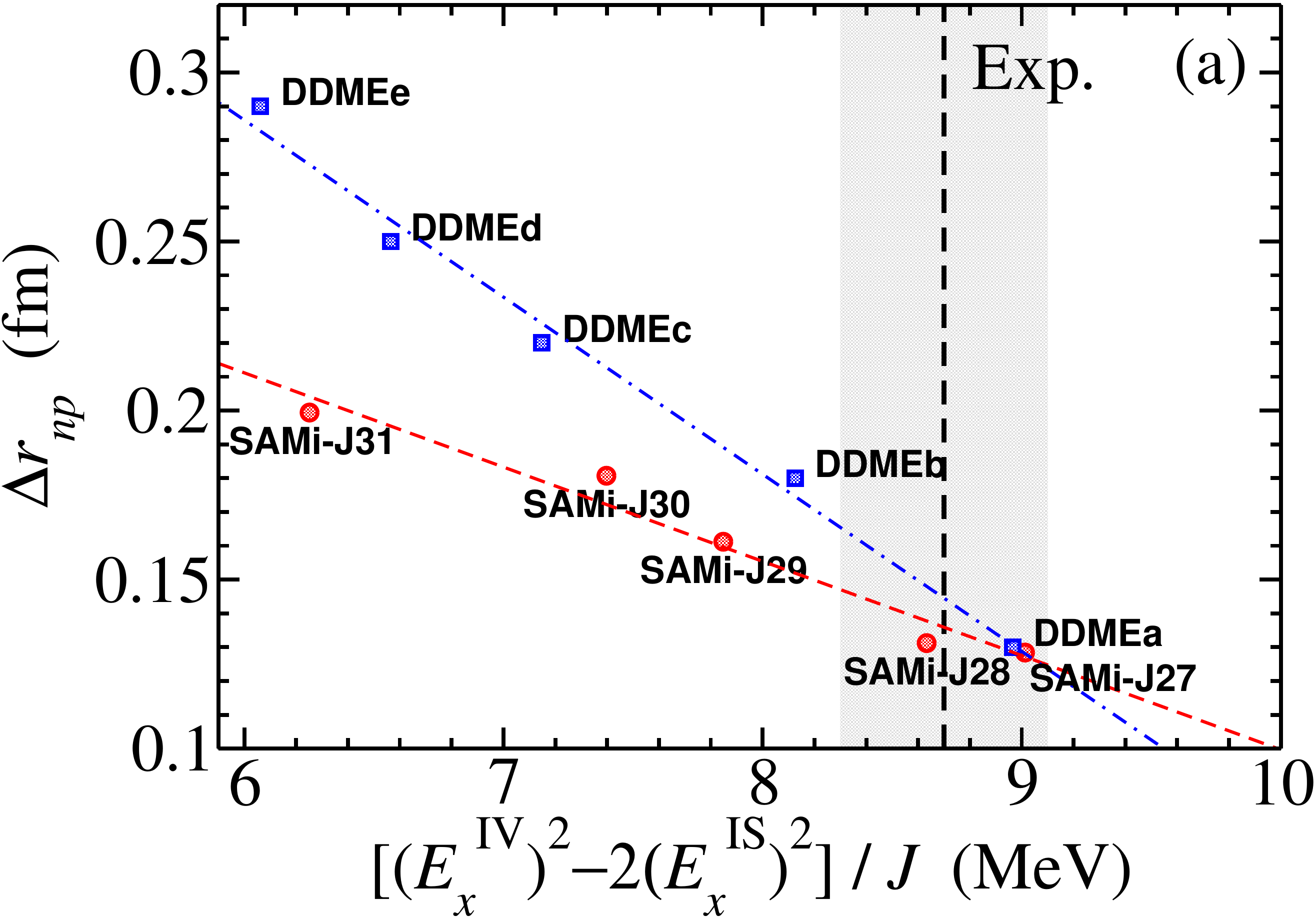}\hfill
\includegraphics[width=0.45\linewidth,clip=true]{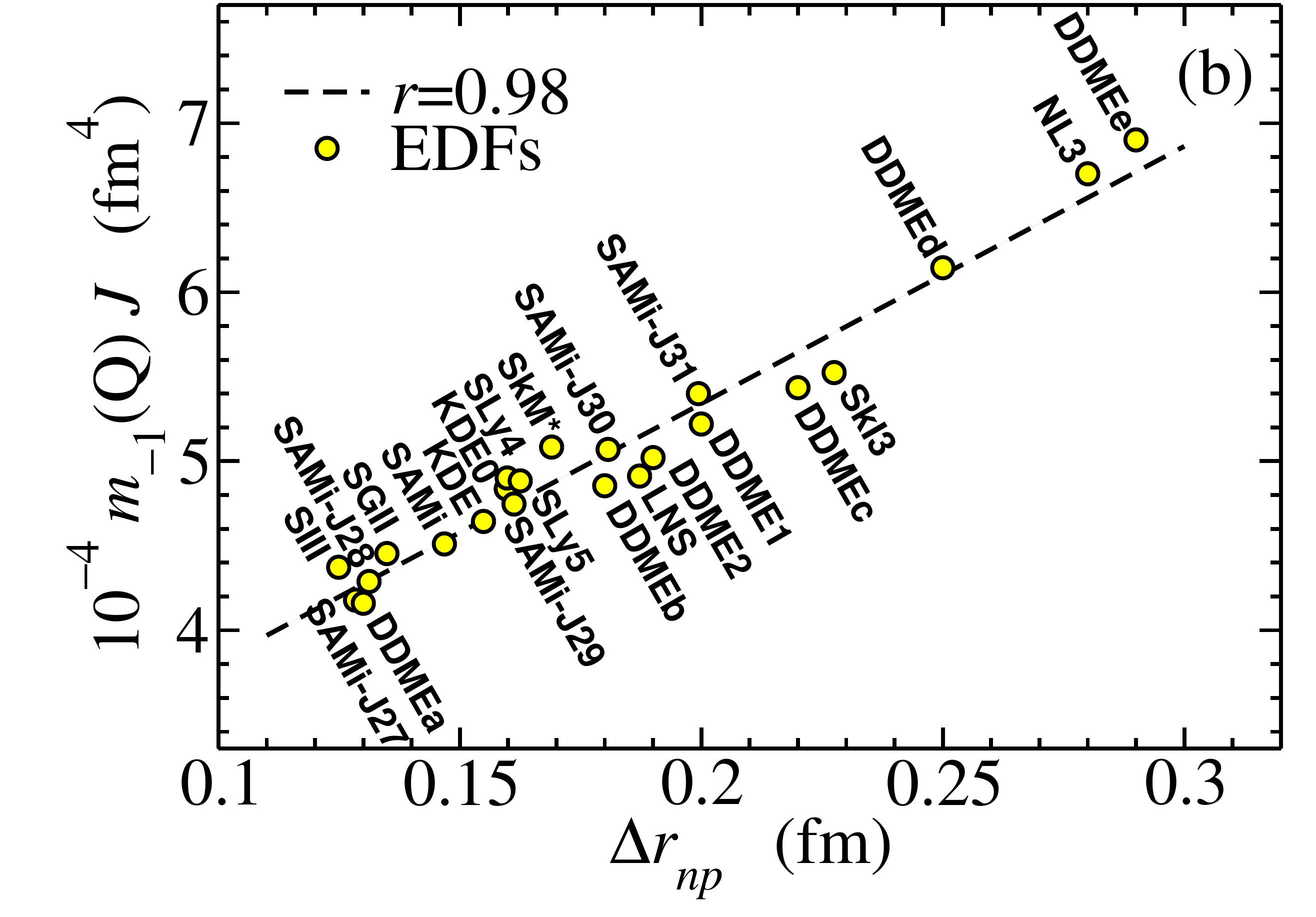}
\caption{Correlations predicted by different EDFs for the case of ${}^{208}$Pb (see text for details). Panel (a): Neutron skin thickness as a function of a combination of the excitation energies of the IS and IV GQRs and the symmetry energy at saturation. We also depict the experimental region as a vertical band (figure adapted from Ref.~\cite{roca-maza13a}). Panel (b): Inverse energy weighted moment of the IV quadrupole response function multiplied by the symmetry energy at saturation as a function of the neutron skin thickness. Correlation coefficient $r$ is also shown.}
\label{fig1}
\end{figure}

In this section, we briefly present part of our results on the analysis of the excitation energies of the isoscalar and isovector GQRs in ${}^{208}$Pb \cite{roca-maza13a}: $E_x^{\rm IS}$ and $E_x^{\rm IV}$, respectively; and an unpublished study on the inverse energy weighted moment of the IV quadrupole response [$m_{-1}(Q)$] also in ${}^{208}$Pb, both based on the predictions of different Energy Density Functionals (EDFs). 

Based on a Quantum Harmonic Oscillator (QHO) approach \cite{bohr69} and the expression for the neutron skin thickness $\Delta r_{np}\equiv \langle r_n^2\rangle^{1/2} - \langle r_p^2\rangle^{1/2}$ of a heavy nucleus in the Droplet Model (DM) \cite{myers74}, one finds an analytical expression that explicitly relates $\Delta r_{np}$ with a combination of the IS and IV excitation energies, $\left[\left(E_x^{\rm IV}\right)^2 - 2\left(E_x^{\rm IS}\right)^2\right]$, and the symmetry energy at nuclear saturation $S(\rho_\infty)\equiv J$ \cite{roca-maza13a}. This correlation is also predicted by more fundamental calculations as it can be seen in Fig.~\ref{fig1}(a) where two families of systematically varied EDFs are shown. We have considered a non-relativistic (SAMi \cite{roca-maza13a}) and a relativistic (DD-ME \cite{vretenar03}) families in which $J$ was systematically varied from 27 MeV to 31 MeV in steps of 1 MeV and 30 MeV to 38 MeV in steps of 2 MeV, respectively. This kind of analysis allows to identify possible correlations between isovector properties. An experimental region is also depicted in Fig.~\ref{fig1}(a) using a reasonable value of $J$ = 32 $\pm$ 1 MeV, in agreement with available estimates \cite{tsang12}, and the experimental weighted average on $E_x^{\rm IV}=22.7\pm 0.2$ MeV  and $E_x^{\rm IS}=10.9\pm 0.1$ MeV \cite{roca-maza13a}. 

Within the same macroscopic approach but this time considering the approximate equality between the DM parameter characterizing the symmetry energy of ${}^{208}$Pb, $a_{\rm sym}(A=208)$, and the symmetry energy of the infinite system, $S(\rho=0.1$ fm${}^{-3})$ ---satisfied to a very good accuracy in EDFs \cite{centelles09}, one finds the following expression for the potential contribution to the symmetry energy \cite{roca-maza13a}, 
\begin{equation}
\vspace{-1mm}
S^{\rm pot}(\rho=0.1{\rm fm}{}^{-3}) = 
\frac{A^{2/3}}{24\varepsilon_{{\rm F}_{\infty}}}
\left[\left(E_x^{\rm IV}\right)^2 - 2\left(E_x^{\rm IS}\right)^2\right] 
\approx 
\frac{1}{25\;{\rm [MeV]}}
\left[\left(E_x^{\rm IV}\right)^2 - 2\left(E_x^{\rm IS}\right)^2\right] \;,
\label{eq1}
\vspace{-1mm}
\end{equation}
as a function of the excitation energies of the IS and IV GQRs and the Fermi energy at saturation density: $\varepsilon_{{\rm F}_\infty}\approx 36$ MeV. Note that the kinetic contribution to the symmetry energy, $S^{\rm kin}(\rho)$, is known to be around $\varepsilon_{{\rm F}_\infty}/3$ for non-relativistic EDFs and around $\varepsilon_{{\rm F}_\infty}/2$ for relativistic EDFs (see Table II of Ref.~\cite{roca-maza13a}).  By inserting the weighted averages of the experimental values, we find $S^{\rm pot}(0.1$ fm${}^{-3}) = 11.1 \pm 0.3$ MeV, in very good agreement with the estimate reported in Ref.~\cite{trippa08} if we consider a non-relativistic kinetic contribution that is consistent with the QHO approach. 

The inverse energy weighted moment associated to the isovector quadrupole response in nuclei, proportional to the quadrupole polarizability, can be derived via a constrained calculation assuming the DM approach, in analogy with the dipole case \cite{roca-maza13b}. Such a result can be expressed in terms of DM parameters and the expression for the neutron skin thickness within the same approach as follows, 
\begin{equation}
\vspace{-1mm}
m_{-1}(Q)\approx\frac{A \langle r^4\rangle}{16\pi J}\left[1+\frac{7}{2}\frac{\Delta r_{np}+
\sqrt{\frac{3}{5}}\frac{e^2 Z}{70 J}-\Delta r_{np}^{\rm surface}}{\langle r^2\rangle^{1/2}(I-I_C)}\right]\;,
\label{iewm}
\vspace{-1mm}
\end{equation}
where $\langle r^2\rangle$ ($\langle r^4\rangle$) is the matter mean square (fourth) radius, $I\equiv (N-Z)/A$, $I_C= (e^2 Z )/(20 J R)$ and $\Delta r_{np}^{\rm surf}$ within the DM is equal to $\sqrt{3/5}[5(b_n^2-b_p^2)/(2R)]$, a correction caused by the difference in the surface widths $b_n$ and $b_p$ of the neutron and proton density profiles. This surface contribution to the neutron skin was shown to be almost constant ($0.09 \pm 0.01$ fm) for the case of ${}^{208}$Pb when predicted by a large set of EDFs ~\cite{centelles10}. As a first reasonable approximation, one can neglect the small variations of $I_C$ and $(e^2 Z)/(70 J)$ and explicitely show within this macroscopic approach that for ${}^{208}$Pb, $m_{-1}(Q) J$ is linearly correlated with $\Delta r_{np}$. Such a correlation is supported by more fundamental microscopic calculations as it can be seen in Fig.~\ref{fig1}(b). A part from the two families of systematically varied EDFs shown in Fig.~\ref{fig1}(a), we also provide in Fig.~\ref{fig1}(b) the predictions of other successful models yielding a correlation coefficient between the discussed quantities of $r=0.98$ (original references for the employed interactions can be found in \cite{roca-maza13a,piekarewicz12}).  

\section{Parity Violating Electron Scattering}
\label{pves}

The feasibility of measuring parity violating electron scattering (PVES) observables in heavy nuclei has been recently demostrated at the Jefferson Laboratory by the PREx collaboration \cite{prex}. In this processes, the parity violating asymmetry $A_{pv}$, very sensitive to the weak charge distribution in nuclei, is measured. This observable isolates the parity violating effect in PVES arising form the fact that electrons with positive (negative) helicity states interact with the Coulomb potential plus (minus) the weak potential created by the target nucleus (see Refs. \cite{donnelly89,horowitz01c,roca-maza11} for further details). 

\begin{figure}
\centering
\sidecaption
\includegraphics[width=0.60\linewidth,clip=true]{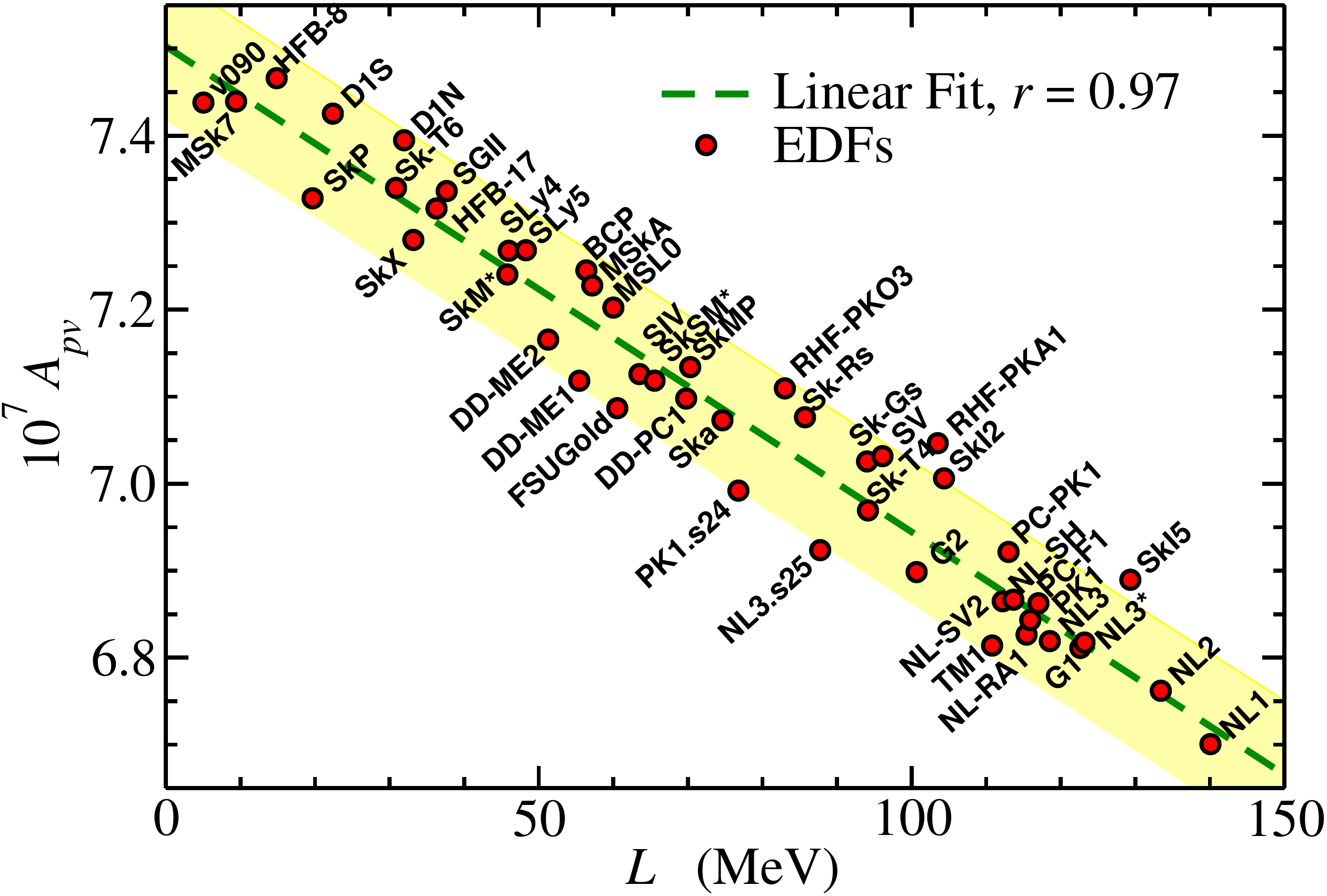}
\caption{Parity violating asymmetry at the PREx kinematics as a function of the $L$ parameter as predicted by different EDFs. A linear fit with correlation coefficient ($r=0.97$) and a 99\% confidence band are also shown (figure adapted from Ref.~\cite{roca-maza13a}).}
\label{fig2} 
\end{figure}

Following a simple Plane Wave Born Approximation (PWBA), one may estimate at a fixed low momentum transfer $q$ that the behaviour of $A_{pv}^{\rm PWBA}$ is dominated by the difference $F_n(q)-F_p(q)\approx q^2\Delta r_{np} \langle r_p^2\rangle^{1/2}/3$ and, thus, by the neutron skin thickness \cite{roca-maza11,centelles10} ---or $L\equiv 3\rho_\infty \partial_\rho S(\rho)\vert_{\rho=\rho_{\infty}}$ if one considers the well known linear correlation between $\Delta r_{np}$ and $L$ within the realm of EDFs \cite{brown00,centelles09}. We show in Fig.~\ref{fig2} the predictions of a large set of EDFs for $A_{pv}$, accounting for Coulomb distortions, at the PREx kinematics \cite{prex} as a function of $L$. We address the reader to Ref.~\cite{roca-maza11} for further details on the theoretical calculations. The correlation found between $A_{pv}$ and $L$ is remarkable ($r=0.97$).    

\section{Conclusions}

Guided by macroscopic models, we have shown that (i) a combination of the $E_x$ of the IS and IV GQRs and $J$ is correlated with $\Delta r_{np}$ of ${}^{208}$Pb within two families of systematically varied EDFs; (ii) the $m_{-1}(Q)$ in ${}^{208}$Pb times $J$ is correlated with $\Delta r_{np}$ of the same nucleus in EDFs; and (iii) that the $A_{pv}$ at the kinematics of PREx is also linearly correlated with $L$ in EDFs. These correlations might be instrumental for the analysis of past and future experiments that aim to improve our knowledge on the isovector sector of the nuclear effective interaction. Finally, we have also estimated the value of the potential energy contribution to the symmetry energy at 0.1 fm$^{-3}$ to be $11.1 \pm 0.3$ MeV using the experimental data on the IS and IV GQRs in the analytical expression derived within QHO.

\begin{acknowledgement}
M.C. and X.~V. acknowledge support from KJCX2-EW-N01; CPAN CSD2007-00042 Programme and Grants No. FIS2011-24154 (MICINN) and No. 2009SGR-1289 (GC); L.~C. from Grants Nos 10875150, 11175216 of the NNSFC and KJCX2-EW-N01 of the PKIP (China); and M.~W. from Grant No. DEC-2011/01/B/ST2/03667 of the NSC (Poland). 
\end{acknowledgement}

%

\bibliography{bibliography}

\begin{thebibliography}{16}

\bibitem{roca-maza13a}
X.~Roca-Maza~{\it et al.}, Phys. Rev. C \textbf{87}, 034301 (2013)

\bibitem{henshaw11}
S.S. Henshaw~{\it et al.}, Phys. Rev. Lett. \textbf{107}, 222501 (2011)

\bibitem{prex}
S.~Abrahamyan~{\it et al.} (PREx Collaboration), Phys. Rev. Lett. \textbf{108},
  112502 (2012)

\bibitem{roca-maza11}
X.~Roca-Maza, M.~Centelles, X.~Vi\~nas, M.~Warda, Phys. Rev. Lett.
  \textbf{106}, 252501 (2011)

\bibitem{brown00}
B.~Alex~Brown, Phys. Rev. Lett. \textbf{85}, 5296 (2000)

\bibitem{tsang12}
M.B. Tsang~{\it et al.}, Phys. Rev. C \textbf{86}, 015803 (2012)

\bibitem{centelles09}
M.~Centelles, X.~Roca-Maza, X.~Vi\~nas, M.~Warda, Phys. Rev. Lett.
  \textbf{102}, 122502 (2009)

\bibitem{trippa08}
L.~Trippa, G.~Col\`o, E.~Vigezzi, Phys. Rev. C \textbf{77}, 061304 (2008)

\bibitem{donnelly89}
T.~Donnelly, J.~Dubach, I.~Sick, Nuclear Physics A \textbf{503}, 589  (1989)

\bibitem{horowitz01c}
C.J. Horowitz, S.J. Pollock, P.A. Souder, R.~Michaels, Phys. Rev. C
  \textbf{63}, 025501 (2001)

\bibitem{bohr69}
A.~Bohr, B.R. Mottelson, \emph{Nuclear Stucture} (W. A.Benjamin Inc., 1975),
  Vol. I \& II

\bibitem{myers74}
W.~Myers, W.~Swiatecki, Ann. Phys. \textbf{84}, 186  (1974)

\bibitem{vretenar03}
D.~Vretenar, T.~Nik\ifmmode \check{s}\else \v{s}\fi{}i\ifmmode~\acute{c}\else
  \'{c}\fi{}, P.~Ring, Phys. Rev. C \textbf{68}, 024310 (2003)

\bibitem{roca-maza13b}
X.~Roca-Maza~{\it et al.}, In preparation  (2013)

\bibitem{centelles10}
M.~Centelles, X.~Roca-Maza, X.~Vi\~nas, M.~Warda, Phys. Rev. C \textbf{82},
  054314 (2010)

\bibitem{piekarewicz12}
J.~Piekarewicz~{\it et al.}, Phys. Rev. C \textbf{85}, 041302 (2012)

\end{thebibliography}

\end{document}